\documentclass[5p,twocolumn]{elsarticle}
\usepackage{graphicx,latexsym}
\usepackage{dcolumn}
\usepackage{amssymb,amsmath,bm}
\usepackage{subfigure}
\usepackage{braket}
\usepackage{siunitx}
\usepackage{array}
\usepackage{float}
\usepackage[nopar]{lipsum}
\usepackage{caption}
\usepackage{multirow}
\usepackage{bigstrut}

\biboptions{sort&compress}

\usepackage[super]{nth}

\usepackage{hyperref}
\hypersetup{
    pdfnewwindow=true,       
    colorlinks=true,         
    linkcolor=blue,          
    citecolor=blue,          
    filecolor=magenta,       
    urlcolor=black           
}

\usepackage[normalem]{ulem}

\def\sec#1{Sec.\ \ref{#1}}
\def\eq#1{Eq.\ (\ref{#1})}
\def\fig#1{Fig.\ \ref{#1}}

\journal{}

\begin{document}

\begin{frontmatter}


\title{Effects of coupling strength of the electron-photon and the photon-environment interactions on the electron transport through multiple-resonances of\break a double quantum dot system in a photon cavity}

\author[a1,a2]{Halo Anwar Abdulkhalaq}
\ead{halo.abdulkhalaq@univsul.edu.iq}
\address[a1]{Division of Computational Nanoscience, Physics Department, College of Science,
	University of Sulaimani, Sulaimani 46001, Kurdistan Region, Iraq}
\address[a2]{Computer Engineering Department, College of Engineering, Komar University of Science and Technology, Sulaimani 46001, Kurdistan Region, Iraq}

\author[a1,a2]{Nzar Rauf Abdullah}
\ead{nzar.r.abdullah@gmail.com}

\author[a4]{Vidar Gudmundsson}
\address[a4]{Science Institute, University of Iceland, Dunhaga 3, IS-107 Reykjavik, Iceland}
\ead{vidar@hi.is}


\begin{abstract}

We study electron transport properties through a double quantum dot (DQD) system coupled to a single mode photon cavity, DQD-cavity. The DQD system has a complex multilevel energy spectrum, in which by tuning the photon energy several anti-crossings between the electron states of the DQD system and photon dressed states are produced, which have not been seen in a simple two level DQD system. Three different regions of the photon energy are studied based on anti-crossings, where the photon energy ranges are classified as ``low", ``intermediate'', and ``high''.
The anti-crossings represent multiple Rabi-resonances, which lead to a current dip in the electron transport at the ``intermediate'' photon energy. Increasing the electron-photon coupling strength, $g_\gamma$, the photon exchanges between the anti-crossing states are changed leading to a dislocation of the multiple Rabi resonance states. Consequently, the current dip at the intermediate photon energy
is further reduced.
Additionally, we tune the cavity-environment coupling, $\kappa$, to see how the transport properties in the strong coupling regime, g$_{\gamma}>\kappa$, are changed for different directions of the photon polarization.
Increasing $\kappa$ with a constant value of $g_\gamma$, a current enhancement in the intermediate photon energy is found, and a reduction in the current is seen for the ``high'' photon energy range.
The current enhancement in the intermediate photon energy is caused by the weakening of the multiple Rabi-resonance in the system.

\end{abstract}


\end{frontmatter}

\section{Introduction}

Quantum dots (QD) are viewed as artificial atoms for which quantum physics play an important role to study their physical properties including the electron-electron and the photon-electron interactions \cite{Kouwenhoven_2001, ashoori1996electrons, PIROT2022413646}.
The QD systems can be coupled to other electron reservoirs via a tunneling regions or contacts that provide possibilities to study quantum transport \cite{Lu2003, ABDULLAH2020114221}. Due to the exhibition of quantum effects in the electronic and the optical properties, in many QD systems the emerging shell structure, leads to photoluminescence with tunable wavelength and variable intensity \cite{acsanm.9b00456, acsanm.0c01386}. QDs have been used in variety of applications ranging from biological \cite{jamieson2007biological} to optoelectronic devices, such as light emitting devices (LED) \cite{shirasaki2013emergence}, and photodetectors \cite{konstantatos2006ultrasensitive, ABDULQADIR2022413907}.

This variety of applications opens doors for further studies of the quantum transport in QD-system coupled to different QED-cavity types such as micro or nano-cavities, in which different parameter, such as the cavity-environment and the electron-photon couplings may play important roles.
The transport properties of QD systems can be controlled by several parameters that can be related to the electron reservoirs, photon cavity and the coupling strengths of the cavity-environment and the electron-photon interaction. The electron transport from the reservoir to the QD system can happen due to the required energy provided by a voltage bias and it is strongly influenced by the single electron charge quantization and the Coulomb interaction \cite{PhysRevB.53.12625}.

Recently, many researcher have focused their attention on the use of quantum dots systems in single photon mode cavities coupled to the environment for the enhancement of spontaneous emission \cite{nano9050671}. Moreover, phenomena such as the Purcell effect in the nanoscale range have been investigated using a quantized photon cavity coupled to an electronic system \cite{Romeira_2018, PurcellEffect2019}. Many projects have studied the use of a cavity photon field for Rabi-resonant states causing oscillations in the electron transport \cite{PhysRevLett.119.223601,Gudmundsson2015}, photon replica states generating photocurrent \cite{Koeppe_2003}, photon-assisted electron transport \cite{PhysRevLett.107.095301}, and thermoelectric inversion in the steady state regime \cite{PhysRevB.83.085428}. Theoretical investigation of transport properties for different electron-photon cavity coupling strength has been reported in \cite{abdullah2020interplay}.

The effects of cavity-environment coupling on current transport have been one of the interesting topics of research. The transport properties has been studied for different values of the cavity-environment coupling strength in the weak and the strong coupling regimes. In the strong coupling regime, where the electron-photon coupling, g$_{\gamma}$, is greater than the cavity-environment coupling, $\kappa$, the vacuum Rabi splitting is observed for a double quantum dot (DQD) system \cite{bruhat2018circuit}. On the other hand, the spontaneous emission rate and the strong Purcell effect are evident in the weak coupling regime where g{$_{\gamma}$$\leq$} $\kappa$ \cite{ spontaneous-englund2005controlling}. Both coupling regimes serve in varieties of applications. For the weak coupling a high efficiency can be achieved due to the high performance of a single-photo source, which is used as solid state interferometric device \cite{ding2016c}. While for optoelectronic nanodevices, such as solid state quantum processors, the desirable regime is the strong coupling regime \cite{wu2019programmable}.

In the present work we study the transport properties in the strong coupling regime. We model a system composed of a DQD embedded in a short parabolic quantum wire coupled to two electron reservoirs with different chemical potentials. The DQD-wire system is coupled to a rectangular single photon mode cavity, and a weak external static magnetic field is applied to the total system. The transport properties are investigated in the steady state regime using a Markovian version of a master equation for a multi-level interacting system \cite{Efficient342017}. The goal of the study is to show the effect of the electron-photon and the cavity-environment coupling on the transport properties in the strong coupling regime. The total current, the partial current and the partial occupation as a function of photon energy are shown for different values of the electron-photon and the cavity-environment couplings.

The following sections of the paper are organized as: \sec{Model_Theory} is devoted to the description of the model and theoretical formalism. In \sec{results} the obtained results are presented, and conclusions are presented in \sec{conclusion}.

\section{Method and Theory}\label{Model_Theory}

In this section, the Hamiltonian of the total system and the constitutive parts is shown with a brief description of the master equation formalism of the open system influenced by the environment.
A double quantum dot (DQD) embedded in a short quantum wire is considered, which is made of GaAs material with the effective mass of $m^* = 0.067 m_{\rm e}$, where $m_{\rm e}$ is the electron mass. The DQD-wire system is connected to two leads from left and right, and it is coupled to a quantized single mode photon cavity. The leads chemical potentials set a bias window in which the electron states fall inside or outside. The electron states can be shifted into or out of the bias window by a potential created by the plunger gate $V_g$.

The  wire is a quasi-one dimensional system confined parabolically in the $y$-direction, but hard-wall confined in the $x$-direction at $x=\pm L_x/2$, where $L_{x}=150$ nm is the length of the quantum wire.
In the parabolic quantum wire are embedded two symmetrical quantum dots placed asymmetrically, described
by the potentials \cite{2019b}
\begin{equation}
	V_{\rm DQD}=\sum_{i=1}^{2}V_{d}^i
	\exp{\{-\beta_i^2(x-x_{0i})^2-\beta_i^2(y-y_{0i})^2\}},
\end{equation}
where $V_d^1=V_d^2=-4.5$ meV is the depth of both quantum dots, $\beta_1=\beta_2= 0.03$ nm$^{-1}$, $x_{01}=23.802$ nm, $x_{02}=-23.802$ nm, $y_{01}=47.604$ nm, $y_{02}=-47.604$ nm.

\begin{figure}[htb]
	\centering
	\includegraphics[width=0.45\textwidth]{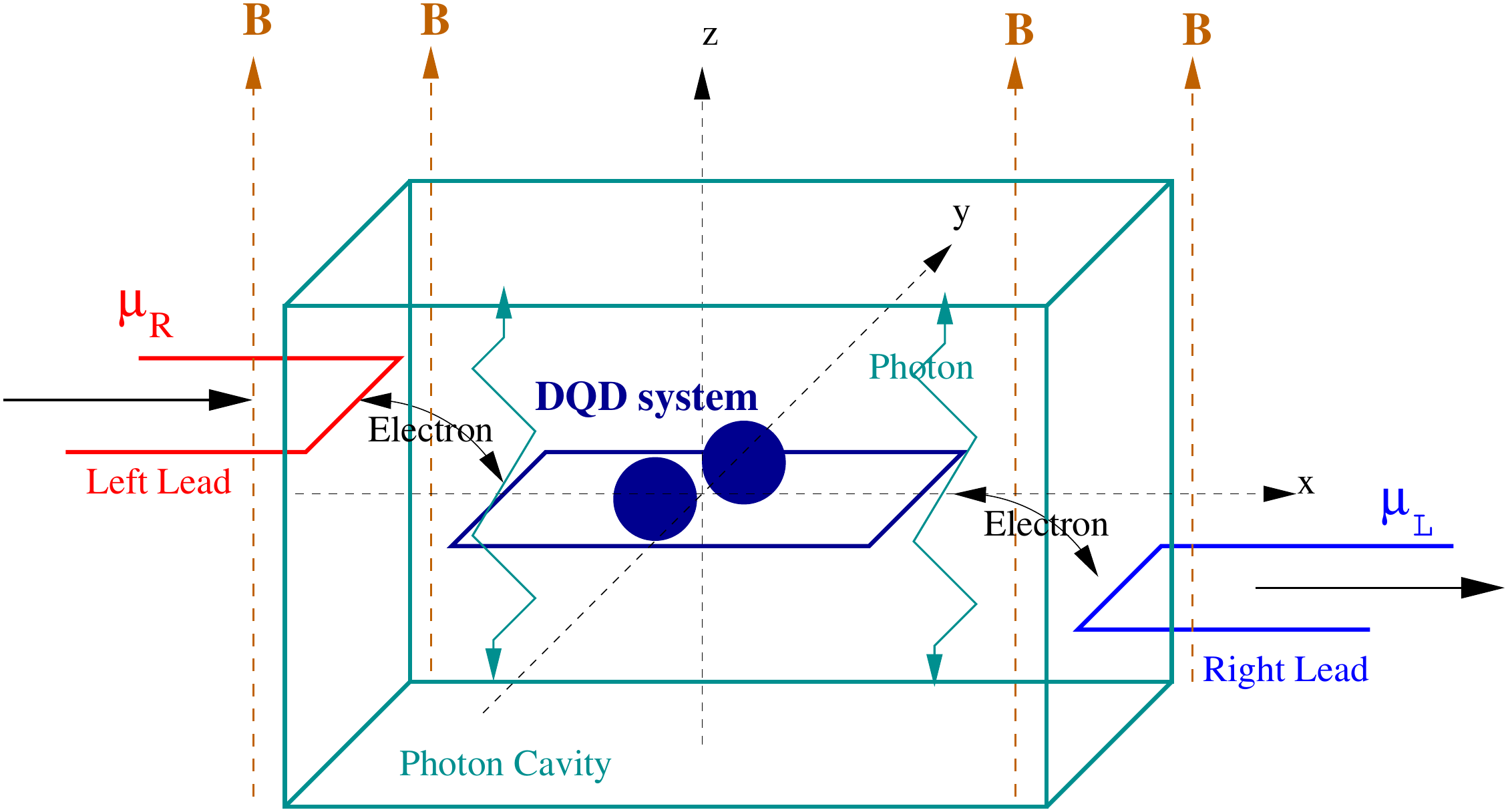}
	\caption{ Schematic diagram showing the DQD-system (dark blue) connected to the left lead (red) and right lead (blue), coupled to a cavity photon field (cyan). The black solid arrows indicate the direction of electron transport in the system and the brown dashed arrows the weak external magnetic field.}
	\label{fig01}
\end{figure}

Initially a weak perpendicular external magnetic field $\textbf{B}=B\hat{\bm z}$ with magnitude of $0.1$ T is applied to the total system, the DQD system and the leads. The effective confinement frequency of the magnetic field is given by $\Omega_w^2=w_c^2+\Omega_0^2$, where the $\omega_c=eB/(m^*c)$ is the cyclotron frequency. A natural length scale of the system is the effective magnetic length, $a_w=\{\hbar/(m^*\Omega_w)\}^{1/2}$. A schematic diagram of the system is displayed in \fig{fig01}.

The many-body (MB) Hamiltonian of the DQD system and the cavity, $H_{\rm S}$, describes the electrons in the central system, controlled by a plunger gate voltage, strongly coupled to a single mode photon field. The electron-electron and the electron-photon interactions are included in the Hamiltonian
\begin{equation}
	H_{\rm S} = H_{\rm e} + H_{\rm \gamma} + H_{\rm e\text{-}\gamma},
	\label{H_S}
\end{equation}
where $H_{\rm e}$ is the DQD system Hamiltonian, $H_{\rm \gamma}$ is the cavity photon field Hamiltonian, and $H_{\rm e\text{-}\gamma}$ describes the interaction of the electrons in the DQD system and the cavity photons. The electronic Hamiltonian of the DQD system is given by \cite{Coexisting2019}
\begin{equation}
	\begin{split}
		H_{\rm e} & =\sum_{nn'}\bra{\Psi_{n'}}\left[\frac{\bm{\pi}_e^2}{2m^*}+eV_g+V_{\rm DQD}\right]\ket{\Psi_n}d_{n'}^{\dagger}d_n \\
		& +H_Z+\frac{1}{2}\sum_{nn'mm'}V_{nn'mm'}d_{n'}^{\dagger}d_{m'}^{\dagger}d_nd_m,
	\end{split}
	\label{H_e}
\end{equation}
where $|\Psi_n\rangle$ is a single-electron eigenstate (SES), and $d_{n'}^{\dagger}$ ($d_n$) are the electron creation (annihilation) operators of the central system. $\bm{\pi}_e=\textbf{p}+\frac{e}{c}\textbf{A}_\mathrm{ext}$, where $\textbf{p}$ is the momentum operator and the vector potential of the external magnetic field \textbf{B} is $\textbf{A}_\mathrm{ext}=-B_y\hat{x}$. The second term of \eq{H_e} is the Zeeman Hamiltonian, $H_Z=\pm g^*\mu_BB/2$, where $g^*$ is the effective Lande $g$-factor and $\mu_B$ is the Bohr magneton. The Coulomb interaction is the third term of \eq{H_e} containing the Coulomb integrals given by
\begin{equation}
	V_{nn'mm'}=\bra{\Psi_{n'}\Psi_{m'}}\left[\frac{e^2}{\bar{\kappa}|\bm{r}-\bm{r}'|}\right]\ket{\Psi_n\Psi_m},
\end{equation}
where $\bar{\kappa}$ is the dielectric constant, and $|\bm{r}-\bm{r}'|$ is the electron pair spatial separation.
The Hamiltonian of free photon field in \eq{H_S} is written as $H_{\rm \gamma} = \hbar \omega_{\rm \gamma} N_{\rm \gamma}$, where $\hbar \omega_{\rm \gamma}$ is the photon energy, and $N_{\rm \gamma} = a^{\dagger}a$ is photon number operator with the photon creation and annihilation operators $a^{\dagger}$ and $a$ respectively. The final term in \eq{H_S} is the Hamiltonian for the electron-photon interaction given by
\begin{equation}
	\begin{split}
		H_{\text e\text{-}\gamma} & =\frac{e}{m^*c} \sum_{nn'}\bra{\Psi_{n'}}{\bm \pi}_e\cdot\textbf{A}_{\gamma}\ket{\Psi_n}d_{n'}^{\dagger}d_n \\
		& +\frac{e^2\textbf{A}_{\rm \gamma}^2}{2m^*c^2}\sum_{nn'}\braket{\Psi_{n'}|\Psi_n}d_{n'}^{\dagger}d_n,
	\end{split}
\end{equation}
with the quantized vector potential of the photon field written as $\textbf{A}_{\rm \gamma}={\cal A}_\gamma (a+a^{\dagger}) \hat{\textbf{e}}$, in which ${\cal A}_\gamma$ is the amplitude of the photon field, $\hat{\textbf{e}}=(e_x,0)$ and $\hat{\textbf{e}}=(0,e_y)$ are the unit vectors of the $x$-polarization and $y$-polarization of the photon field, respectively. The electron-photon coupling strength can be defined from the vector potential amplitude as g$_{\gamma}= {\cal A}_\gamma \Omega_w a_w/c$. The fact that the wavelength of the FIR (far-infrared) cavity field is much larger than the size of the DQD system leads to the simple form for the vector potential $\bm{A}_\gamma$. The numerical exact diagonalization technique, using a tensor product of the Coulomb interacting MB bases and states of the photon number operator, is used for the electron-photon Hamiltonian \cite{2019b,Stepwise442012}.

A Markovian master equation is applied for calculating the electron transport through the DQD system coupled to the leads in the steady state regime. The projection of the total system onto the subspace of the DQD system is the base for the derivation of the master equation. The formalism was derived from the Nakajima and Zwanzig generalized master equation (GME) \cite{nakajima1958quantum, zwanzig1960ensemble}. The reduced density operator kernel of the Nakajima Zwanzig integro-differential equation for the central system is approximated up to second order terms in the system-leads couplings \cite{Coexisting2019}.
In the dynamical regime of our interest, which is steady state regime, a Markovian approximation is used for transforming the GME to the Liouville space of transitions \cite{Efficient342017}. The Markovian master equation describes the dynamical evolution of the reduced density operator $\rho_{\rm S}$, which describes the time evolution of the central system under the influence of the photon reservoir and the external leads \cite{Vidar_2021} as is presented in \cite{correlations45b2018}. The tensor product of the uncorrelated reduced density operators of the central system $\rho_{\rm S}(t_0)$ and external leads $\rho_l(t_0)$, before the coupling of central system to
the external leads, gives the total density operator $\rho(t_0)$ as $\rho(t_0)=\rho_l(t_0)\rho_{\rm S}(t_0)$,
where $t_0$ represents a time before the coupling. After the coupling at $t>t_0$, the reduced density operator of the central system  is written as \cite{Efficient342017}
\begin{equation}
	\rho_{\rm S}(t)=Tr_l[\rho],
\end{equation}
where $l$ represents the the left L and the right R leads of the electron reservoirs.
From obtaining the reduced density operator the current transport through the DQD system can be calculated by
\begin{equation}
	I_{L,R}=Tr[\rho_{\rm S}^{L,R}(t) \, Q]
\end{equation}
where $\rho_S^{L,R}(t)$ is reduced density operator, and $Q=-e\sum_{n}d_n^{\dagger}d_n$ is the charge operator of the DQD system.

\section{Results}\label{results}

The obtained results of our calculations are presented in this section. We investigate the effect of the electron-photon and the cavity-environment coupling strengths on the transport properties of the DQD system. The investigated DQD system is a short quantum wire of length $L=150$~nm embedded with two symmetrical quantum dots, the system is coupled to a single mode photon cavity.
In addition, the system is connected to two leads with chemical potentials $\mu_L=1.25$ meV, and $\mu_R=1.15$ meV, where the temperature of the leads is assumed to be constant at $T=0.5$~K, and the system is placed in a weak external magnetic field B$=0.1$ T.
\begin{figure}[htb]
	\centering
	\includegraphics[width=0.48\textwidth]{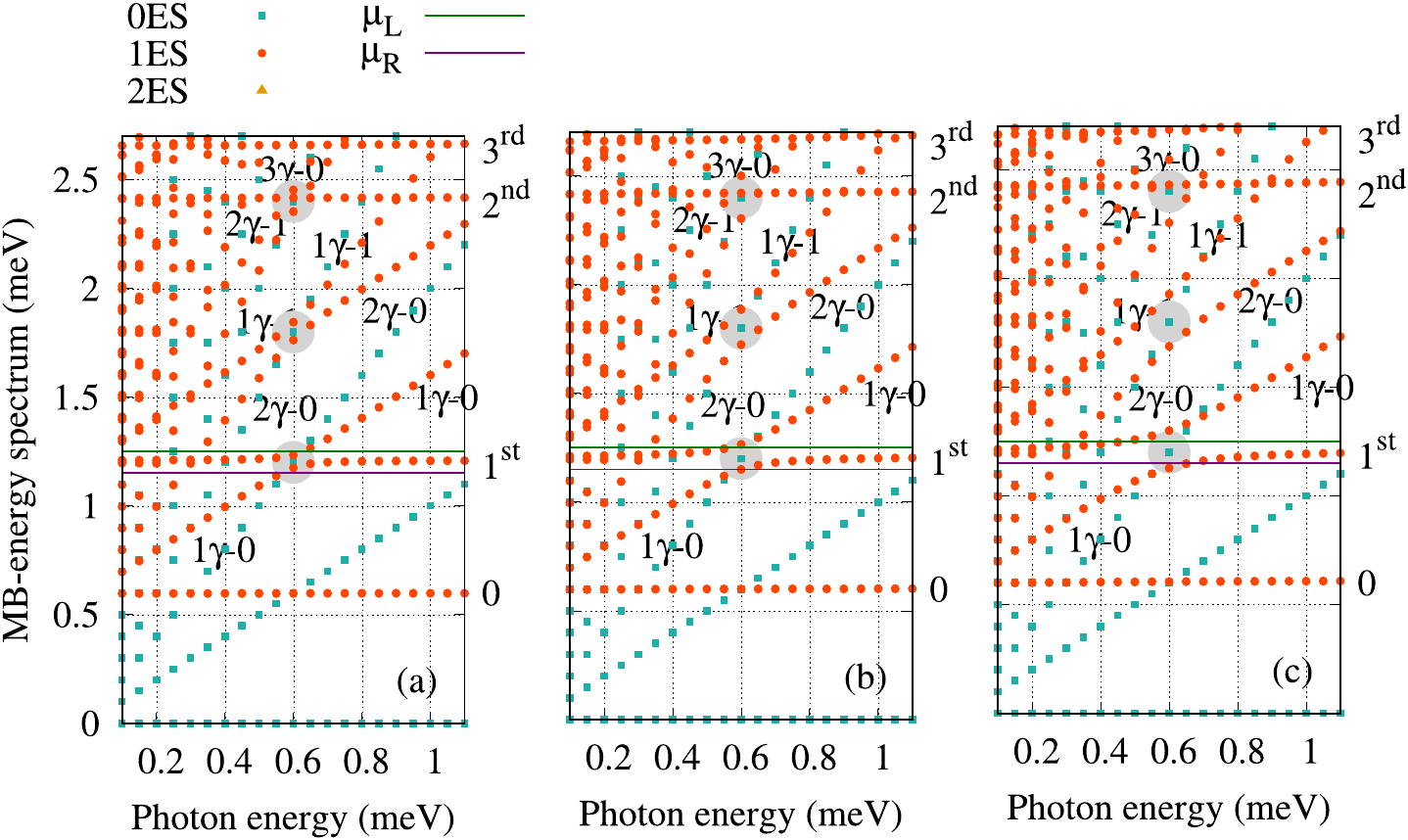}
	\caption{Many-Body (MB) energy spectrum as a function of photon energy for some selected states of a DQD system coupled to a photon cavity for three different values of the electron-photon coupling strength (a) g$_{\gamma}=0.1$, (b) g$_{\gamma}=0.2$ and (c) g$_{\gamma}=0.3$ meV for $x$-polarization, where 0ES (green squares) represent zero-electron states, 1ES (red circles) are one-electron states, and 2ES (orange triangles) are two-electron states. The green and purple lines are the chemical potential of left lead, $\mu_L=1.25$ meV, and right lead, $\mu_R=1.15$ meV, respectively. 0 is one-electron ground-state energy, while 1$^{\rm st}$ , 2$^{\rm nd}$, and 3$^{\rm rd}$ are one-electron first-, second-, and third-excited state respectively. 1$\gamma$-0, 2$\gamma$-0, and 3$\gamma$-0, respectively, refer to the first, the second, and the third photon replica of ground state, while 1$\gamma$-1 and 2$\gamma$-1 are first and the second photon replica of the excited states respectively. The initial cavity photon number is n$_R=1$ and the cavity-environment coupling strength is $\kappa=10^{-5}$ meV. The system is placed in a weak external magnetic field $B=0.1$ T, the plunger gate voltage is $V_g=0.6$ meV, the leads temperature is $T_{L,R}=0.5$ K.}
	\label{fig02}
\end{figure}

The many-body (MB) energy spectrum of the DQD system coupled to cavity as a function of photon energy
is presented in \fig{fig02} for three different values of electron-photon coupling strength g$_{\gamma}=0.1$ (a), $0.2$ (b), and $0.3$~meV (c). The green squares represent zero-electron states (0ES), red circles are one-electron states (1ES), and orange triangles are two-electron states (2ES).
The green and purple lines are the chemical potential of left lead and right lead respectively. An energy range from 0.0 to 2.7 meV is selected such that four lowest energy states are included which are: 0 is one-electron ground-state, while 1$^{\rm st}$ , 2$^{\rm nd}$, and 3$^{\rm rd}$ are one-electron first-, second-, and third-excited state respectively. The 1$\gamma$-0, 2$\gamma$-0, and 3$\gamma$-0, respectively, refer to the first, the second, and the third photon replica of the ground state, and 1$\gamma$-1 and 2$\gamma$-1 are first and second photon replicas of the excited states, respectively, are included in the selected energy range. It can be seen that the original electronic states, such that the first excited state inside the bias window, have an almost flat dispersion throughout all photon energies in the MB-energy spectrum. On the other hand, the location of photon replica states varies with the changing photon energy, that results in strong dispersion for these states.
Anti-crossings are produced in the energy spectrum with the tuning of the photon energy. There are three significant anti-crossings, highlighted with gray circles, formed at photon energy $\hbar \omega_{\rm \gamma}=0.6$ meV. It is expected that with increasing electron-photon coupling strength $g_{\gamma}$ the anti-crossing becomes stronger as can be seen in Fig.\ \ref{fig02} and can be correlated to the photon content of the states.
\begin{figure}[htb]
	\centering
	\includegraphics[width=0.4\textwidth]{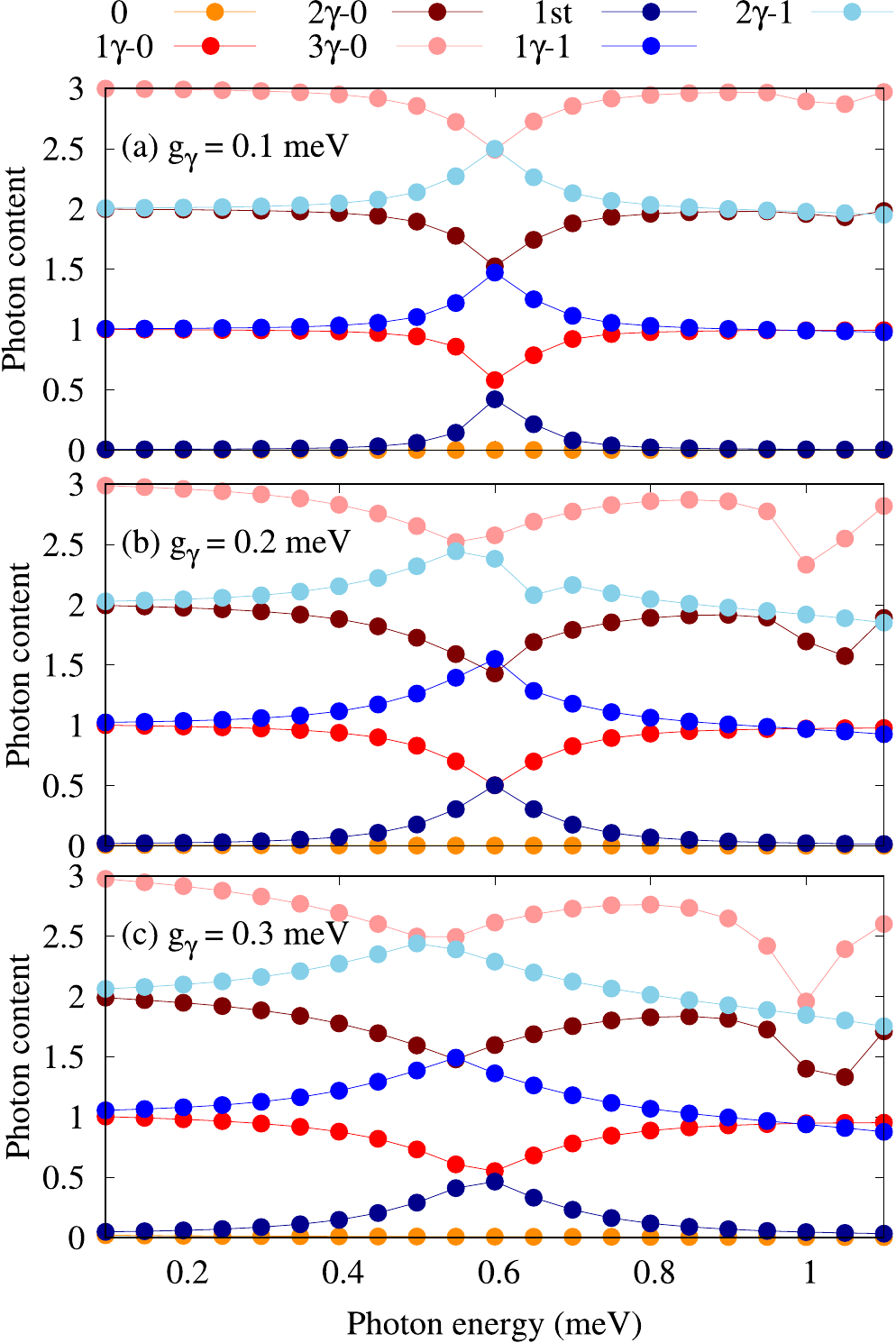}
	\caption{The photon content as a function of photon energy for some selected states of a DQD system coupled to a photon cavity for three different values of electron-photon coupling strength  g$_{\gamma}=0.1$ (a), $0.2$ (b), and $0.3$ meV (c). Th selected states are the ground (orange) and its 1$^{\rm st}$ (red), 2$^{\rm nd}$ (maroon) and 3$^{\rm rd}$ (pink) photon replica states and first excited state (dark blue) and its 1$^{\rm st}$ (blue) and 2$^{\rm nd}$ (cyan) photon replica states. The initial cavity photon number is n$_R=1$ and the cavity-environment coupling strength is $\kappa=10^{-5}$ meV. The system is placed in weak external magnetic field $B=0.1$~T, the plunger gate voltage is $V_g=0.6$ meV, the leads temperature is $T_{L,R}=0.5$ K, and the chemical potential of the left and right leads are $\mu_L=1.25$ and $\mu_R=1.15$ meV respectively.}
	\label{fig03}
\end{figure}

The photon content of the states of the DQD system as a function of the photon energy is shown in \fig{fig03}. The displayed photon content is the content of two lowest selected energy states, the ground, $0$, (orange) and the first-excited, $1^{\rm st}$, (dark blue) states, with their photon replica states generated due to the cavity photon field. The included photon replica states are replicas of ground state 1${\gamma}$-0 (red), 2${\gamma}$-0 (maroon) and 3${\gamma}$-0 (pink), and replicas of the first-excited state 1${\gamma}$-1 (blue) and 2${\gamma}$-1 (cyan) respectively. Throughout the calculations only the $x$-polarization of the photon field is considered, this is because in both $x$- and $y$-directions the photon field has symmetric effect on the DQD system due to the symmetric properties of the quantum dots. It can be seen that tuning the photon energy results in exchange of photons between the selected states, this is a confirmation of the produced anti-crossings caused by the Rabi-resonance between these states.

We observe that the photon content is shared from the photon replica states of ground state to first- excited state and its photon replica states, such that an exchange occurs between 1$^{\rm st}$ (dark blue) and 1$\gamma$-0 (red), 1$\gamma$-1 (blue) and 2$\gamma$-0 (maroon), and 2$\gamma$-1 (cyan) and 3$\gamma$-0 (pink). For all three values of the electron-photon coupling strength, the photon exchange occurs at intermediate range of photon energy from $0.4$ to $0.7$~meV. Maximum photon exchange occurs at 0.6 meV for g$_{\gamma}=0.1$ meV, while a dislocation to lower photon energy can be seen as the electron-photon coupling increases. The dislocation in the photon exchange from $0.6$~meV at higher electron-photon coupling strength indicates a weakening of the multiple Rabi-resonances.

We shall see the effects of these multiple photon exchange and Rabi-resonances on the current, the partial current, and the partial occupation of the DQD system. As mentioned, the photon field polarization has symmetric effect, thus only results for $x$-polarization are presented and the results of $y$-polarization are not shown.

\begin{figure}[htb]
	\centering
	\includegraphics[width=0.45\textwidth]{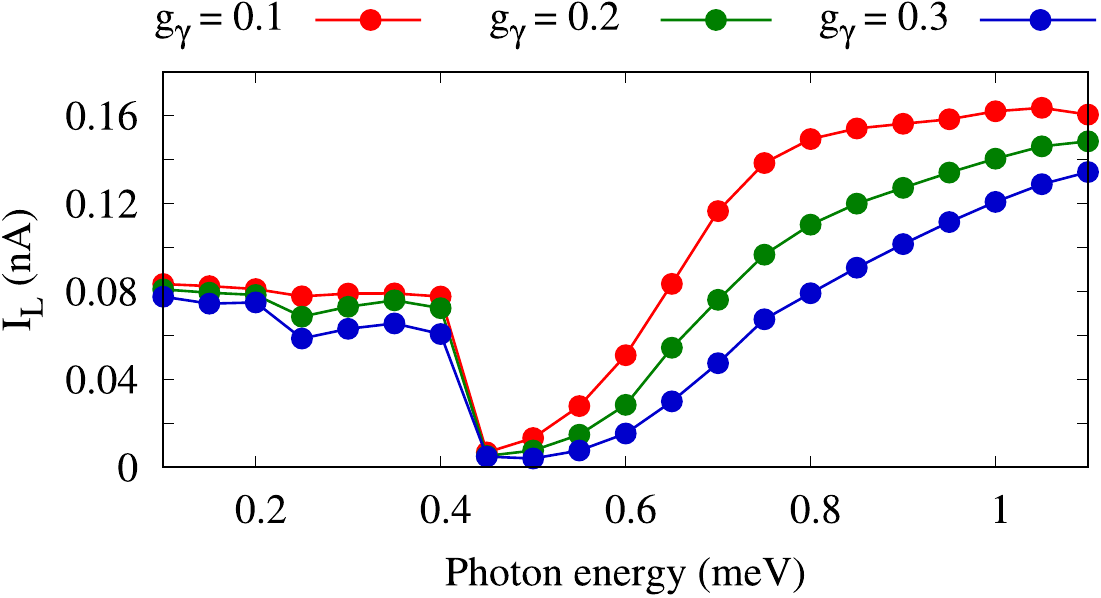}
	\caption{Total current from the left lead to the DQD system as a function of photon energy for $g_{\gamma}=0.1$ (red), 0.2 (green), and 0.3 meV (blue) in case of an $x$-polarized photon field. The initial photon number in the cavity is $n_R=1$, $B=0.1$ T, $\kappa=10^{-5}$ meV, $V_g=0.6$ meV and $T_{L,R}=0.5$ K. }
	\label{fig04}
\end{figure}
In \fig{fig04} the total left current, the current passing through the DQD system from the left lead, as a function of photon energy is shown for $g_{\gamma}=0.1$ (red), 0.2 (green), and 0.3 meV (blue). One can see that there is a current dip in the photon exchange region, $0.4\text{-}0.7$~meV, for all three values of $g_{\gamma}$. The current dip is caused by multiple Rabi-resonances in which the photon exchange plays a role.
We also notice that the total current decreases as the electron-photon coupling increases, more significantly at high values of photon energy, but the current in the dip is almost unchanged in the resonance regions. To understand the nature of the current dip and see which states mostly contribute to the total current, we present the results of the partial currents and the partial occupation of some relevant states of the DQD system.

\begin{figure}[htb]
	\centering
	\includegraphics[width=0.5\textwidth]{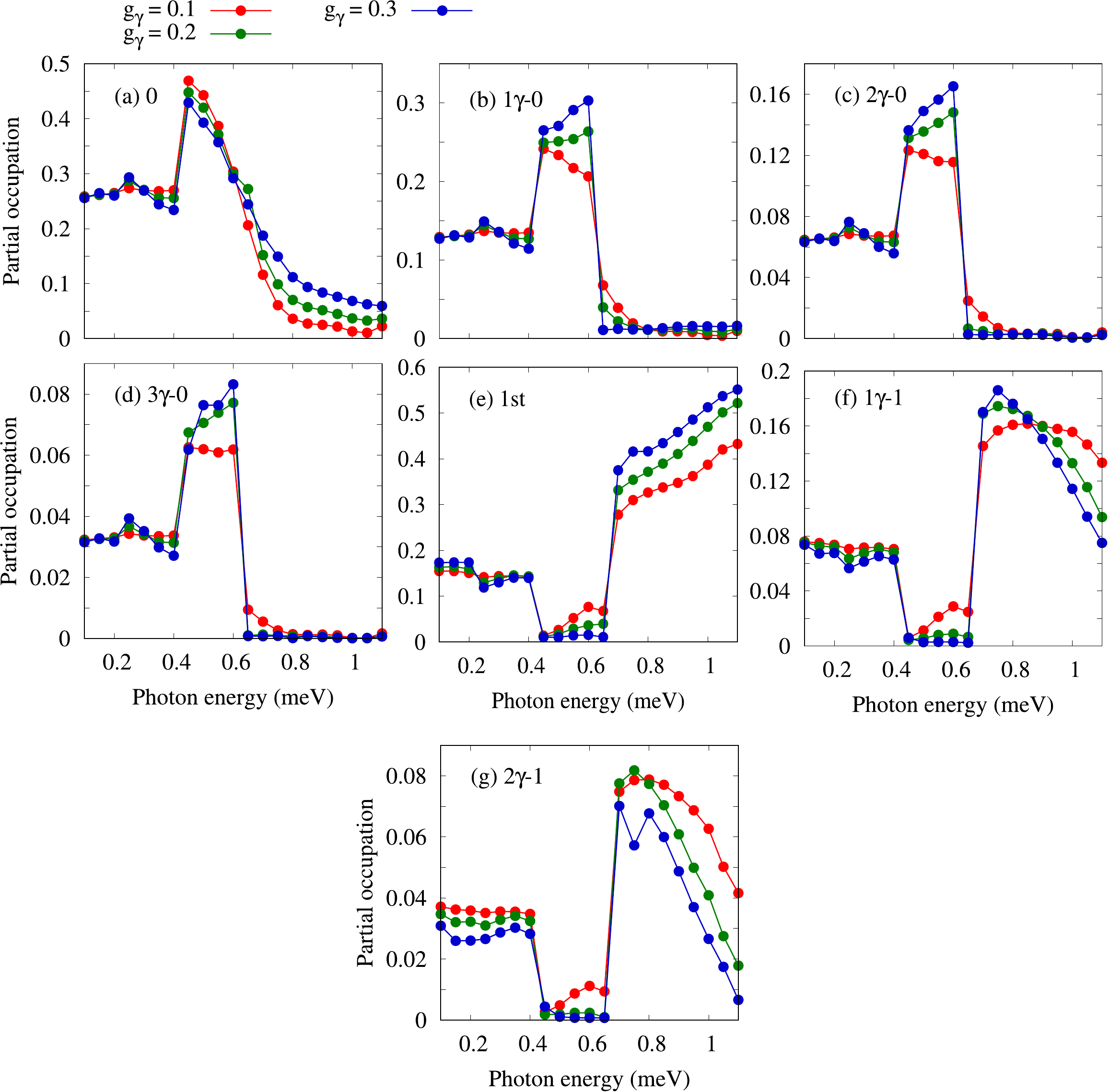}
	\caption{Partial occupation of the DQD system as a function of photon energy for the states (a) 0, (b) 1$\gamma$-0, (c) 2$\gamma$-0, (d) 3$\gamma$-0, (e) 1$^{\rm st}$, (f) 1$\gamma$-1, and (g) 2$\gamma$-1, for three values of electron-photon coupling g$_{\gamma}=0.1$ (red), 0.2 (green), and 0.3 meV (blue) in case of $x$-polarized photon field. The initial photon number in the cavity is $n_R=1$, $B=0.1$ T, $\kappa=10^{-5}$ meV, $V_g=0.6$ meV and $T_{L,R}=0.5$ K.}
	\label{fig05}
\end{figure}
The effect of varying the electron-photon coupling strength, $g_{\gamma}=0.1$ (red), $0.2$ (green)
and $0.3$ meV (blue), on the partial occupation of the DQD system is shown in \fig{fig05}.
In the low photon energy range from $0.1$ to $0.4$~meV a constant occupation is seen for
the selected states for all three values of the electron-photon coupling, while different
characteristics can be seen for the intermediate and the high photon energy ranges.
In the photon exchange region and the intermediate photon energy range from $0.4$ to $0.7$ meV,
there is an increase in the occupation, or the population, for the ground and its photon replica states,
while for the first-excited and its replica states the occupation is decreased.
This increased population and the depopulation is due to the Rabi resonances between the states, as the populated states lose photons and the depopulated state gain photons as is displayed in \fig{fig03}. This population and depopulation of the states is the confirmation of stimulated emission and absorption processes in the resonance region.
It can be seen that the occupation of the ground state \fig{fig05}(a) increases with increasing $g_{\gamma}$ in the photon exchange region, this is also true for the first excited state \fig{fig05}(e) and its photon replica states \fig{fig05}(f) and \fig{fig05}(g). On the other hand, we see that occupation decreases for increasing electron-photon coupling for the photon replica states of the ground state \fig{fig05}(b), \fig{fig05}(c) and \fig{fig05}(d). This is again a confirmation of the Rabi splitting between the states which becomes stronger at higher value of $g_{\gamma}$ (see \fig{fig02}(c)). In this case, the correlation between
the ground state and the first-excited states and their photon replicas gets weaker leading to population of the first-excited state and its replicas and depopulation of the replicas of the ground state in the photon-exchange region when the value of $g_{\gamma}$ is increased. Despite having almost zero photon content and not being in resonance with any other states, the ground state has the highest occupation value in the photon exchange region, this happens due to the stimulated emission-absorption process.

\begin{figure}[htb]
	\centering
	\includegraphics[width=0.5\textwidth]{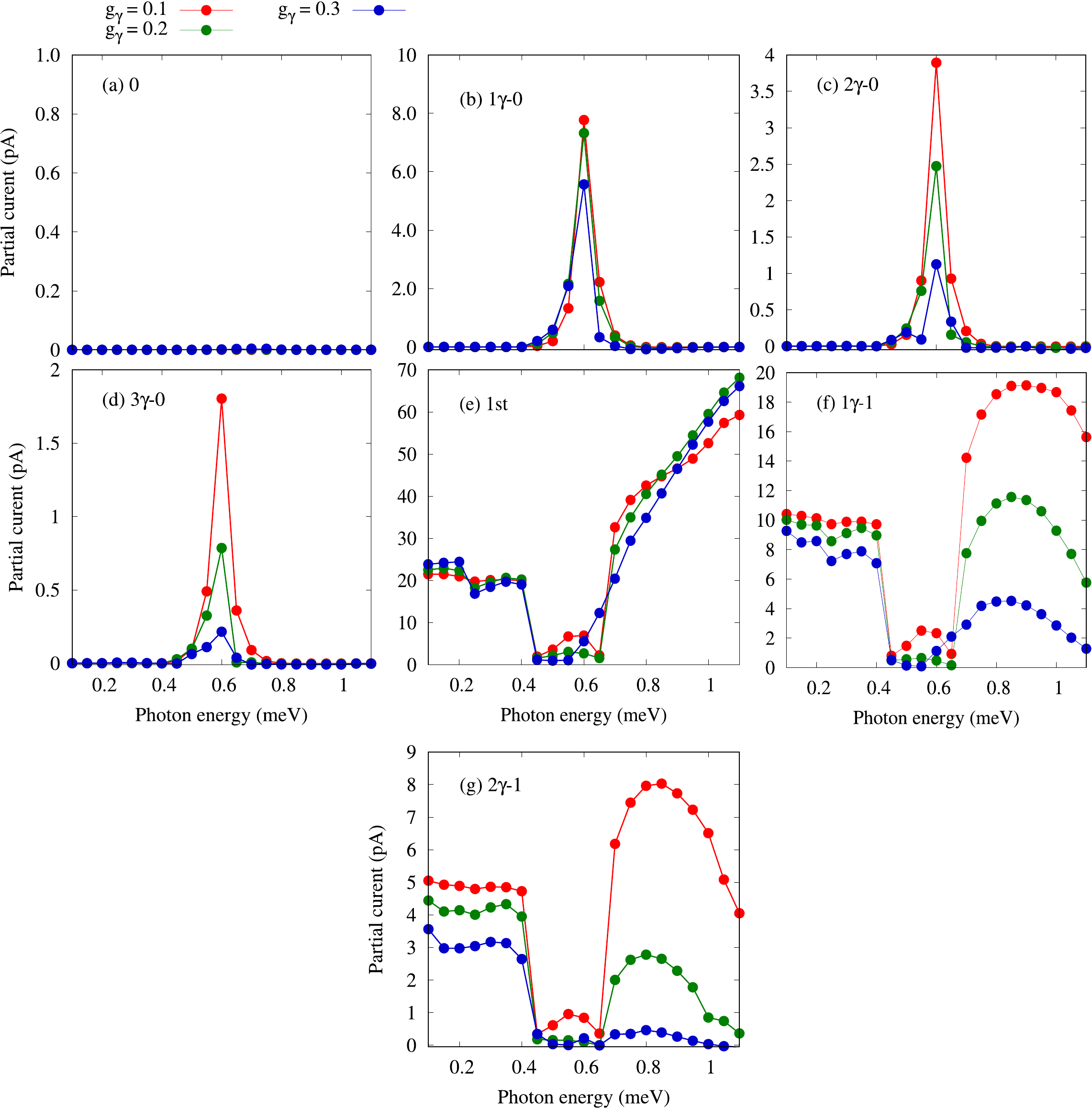}
	\caption{Partial current from the left lead to the DQD system as a function of photon energy for the states (a) 0, (b) 1$\gamma$-0, (c) 2$\gamma$-0, (d) 3$\gamma$-0, (e) 1$^{\rm st}$, (f) 1$\gamma$-1, and (g) 2$\gamma$-1, for three values of electron-photon coupling $g_{\gamma}=0.1$ (red), 0.2 (green), and 0.3 meV (blue) in case of $x$-polarized photon field. The initial photon number in the cavity is $n_R=1$, $B=0.1$ T, $\kappa=10^{-5}$ meV, $V_g=0.6$ meV and $T_{L,R}=0.5$ K.}
	\label{fig06}
\end{figure}
The partial current from the left lead through the DQD as a function of the photon energy is displayed in \fig{fig06} for (a) the ground state, (b) the first-, (c) the second-, (d) the third-photon replica state of ground state, and (e) the first excited state, (f) the first-, (g) the second-photon replica state of the first excited state, for three values of electron-photon coupling $g_{\gamma}=0.1$ (red), 0.2 (green), and 0.3 meV (blue). It can clearly be seen that the current through the ground state, $0$ (a), is zero, this is because the ground state is located below the bias window. In the intermediate photon energy values, the photon exchange region, there is enhancement of current for the photon replica states of the ground state, while there is a current dip for the first-excited state and its photon replica states. This current enhancement and the dip is caused by multiple resonances in the photon exchange region. The characteristics of the partial current of the first-excited state and its photo replica states are similar to that of the total current. This indicates that these states dominantly contribute to the transport. This is because the first excited state is located inside the bias window. Increasing the value of electron-photon coupling leads to a decrease in the current transport through all the selected states.

\begin{figure}[htb]
	\centering
	\includegraphics[width=0.45\textwidth]{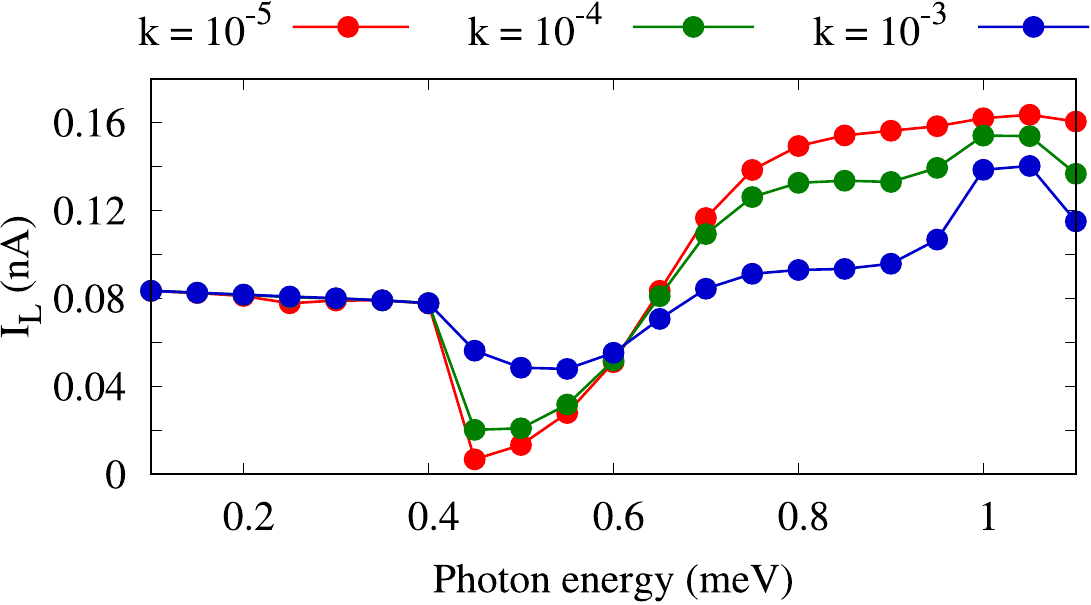}
	\caption{Total current from left lead to the DQD system as a function of photon energy for $\kappa=10^{-5}$ (red), $10^{-4}$ (green), and $10^{-3}$ meV (blue) in case of $x$-polarized photon field. The initial photon number in the cavity is $n_R=1$, $B=0.1$ T, $g_{\gamma}=0.1$ meV, $V_g=0.6$ meV and $T_{L,R}=0.5$ K.}
	\label{fig07}
\end{figure}

\begin{figure}[htb]
  	\centering
  	\includegraphics[width=0.5\textwidth]{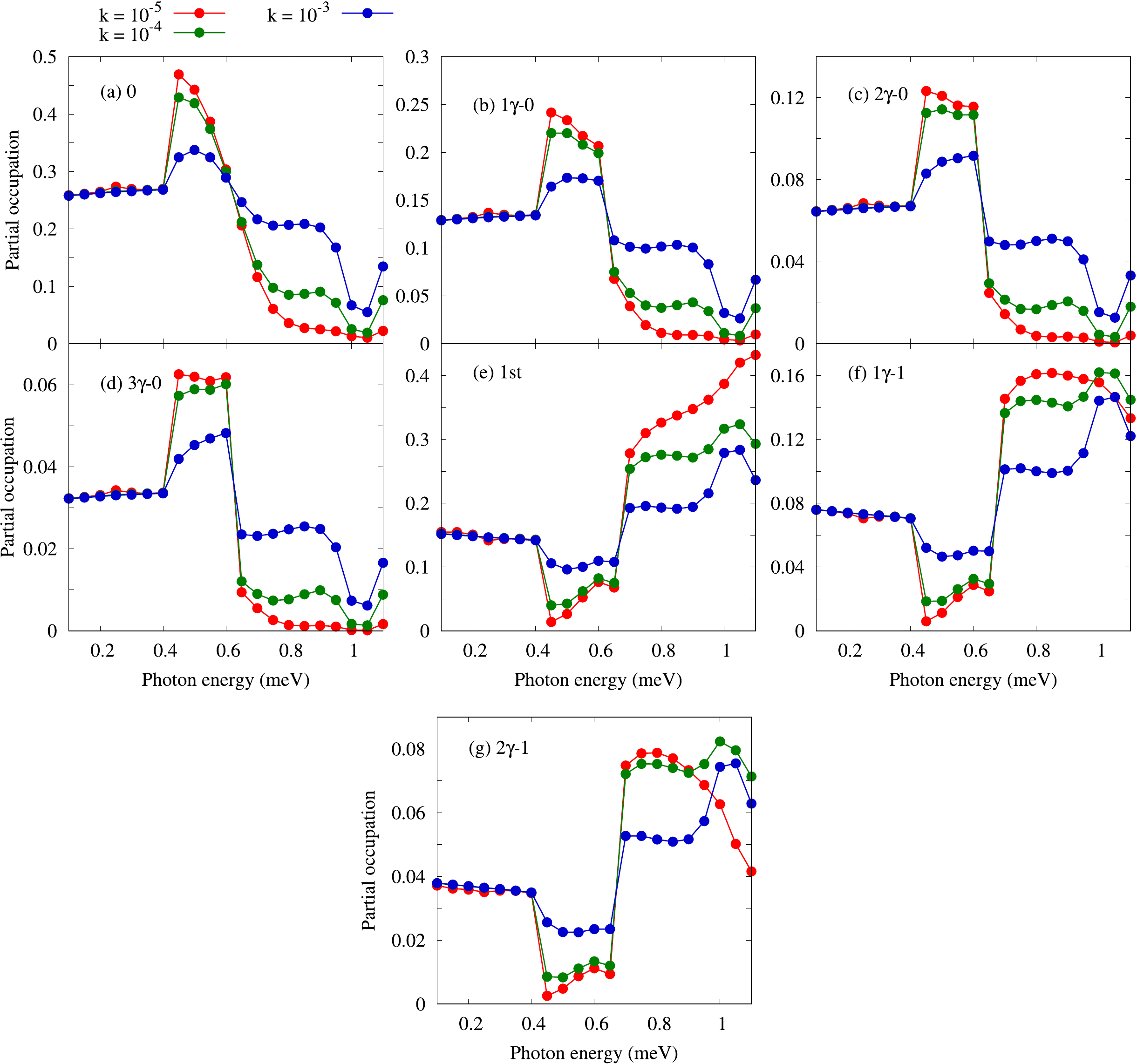}
  	\caption{Partial occupation of the DQD system as a function of photon energy for the states (a) 0, (b) 1$\gamma$-0, (c) 2$\gamma$-0, (d) 3$\gamma$-0, (e) 1$^{\rm st}$, (f) 1$\gamma$-1, and (g) 2$\gamma$-1, for three values of cavity-environment coupling $\kappa=10^{-5}$ (red), $10^{-4}$ (green), and $10^{-3}$ meV (blue) in case of x-polarized photon field. The initial photon number in the cavity is $n_R=1$, $B=0.1$ T, $g_{\gamma}=0.1$ meV, $V_g=0.6$ meV, and $T_{L,R}=0.5$ K.}
  	\label{fig08}
\end{figure}

\begin{figure}[htb]
	\centering
	\includegraphics[width=0.5\textwidth]{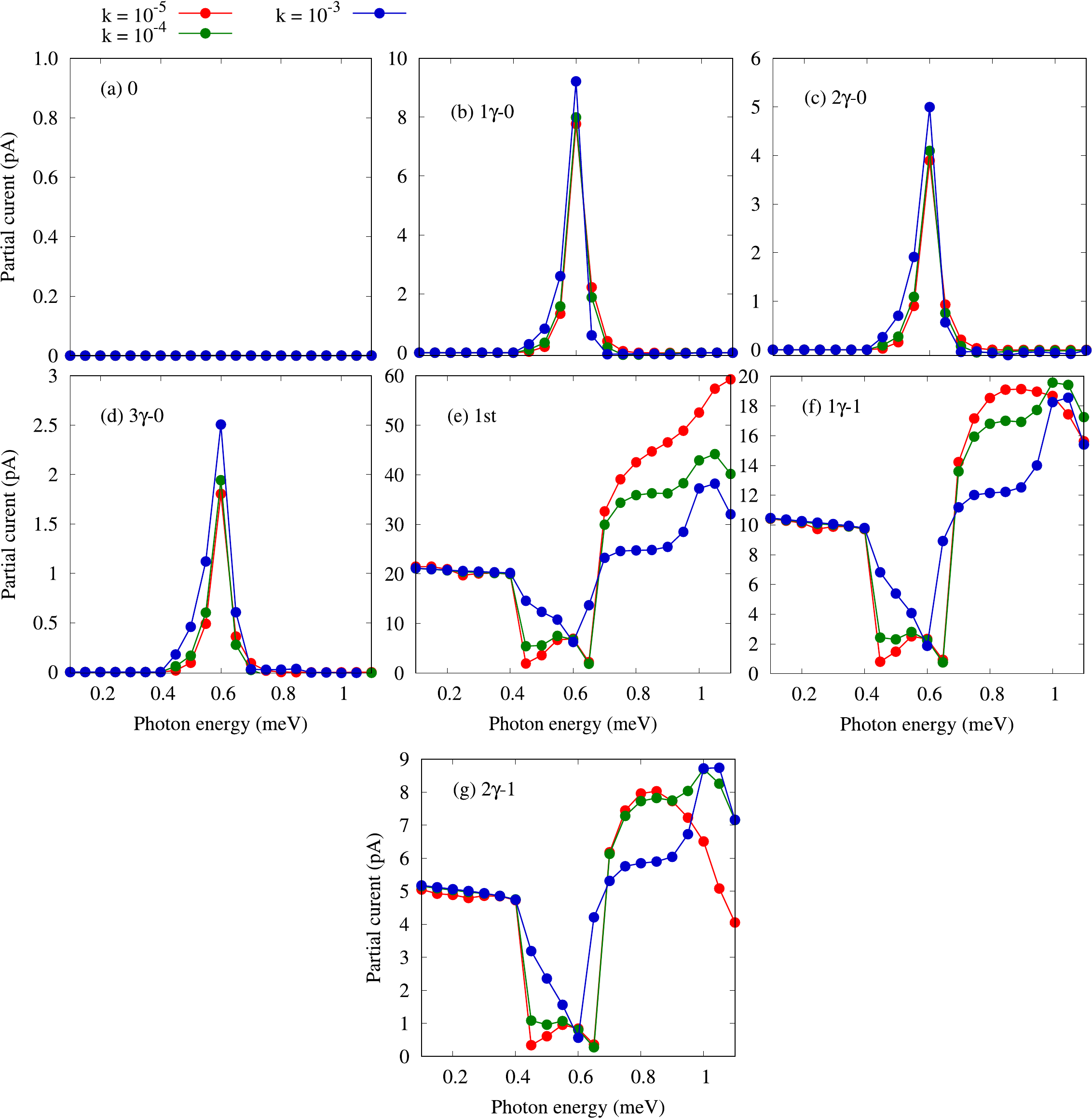}
	\caption{Partial current from the left lead to the DQD system as a function of photon energy for the states (a) 0, (b) 1$\gamma$-0, (c) 2$\gamma$-0, (d) 3$\gamma$-0, (e) 1$^{\rm st}$, (f) 1$\gamma$-1, and (g) 2$\gamma$-1, for three values of the cavity-environment coupling $\kappa=10^{-5}$ (red), $10^{-4}$ (green), and $10^{-3}$ meV (blue) in case of $x$-polarized photon field. The initial photon number in the cavity is $n_R=1$, $B=0.1$ T, $g_{\gamma}=0.1$ meV, $V_g=0.6$ meV, and $T_{L,R}=0.5$ K.}
	\label{fig09}
\end{figure}

Now, we show results for the transport properties of the DQD system, again for the $x$-polarized photon field, when the cavity-environment coupling strength $\kappa$ is varied. Notice we are still in the strong coupling regime, $g_{\gamma}>\kappa$, the value of electron-photon coupling is $g_{\gamma}=0.1$ meV and the initial photon number in the cavity is $n_{\rm R}=1$. The total current from left lead to the DQD system for three values of $\kappa=10^{-5}$ (red), $10^{-4}$ (green), and $10^{-3}$ meV (blue) as a function of the photon energy is shown in \fig{fig07}.
There, can be seen that at low photon energy range, $0.1\text{-}0.4$~meV, the value of the current is similar for all three values of $\kappa$, while in the high photon energy region, $0.7\text{-}1.1$~meV, as $\kappa$ is increased the value of the current is reduced, this in fact corresponds to the tuning of $g_{\gamma}$ in \fig{fig04}. In the intermediate photon energy range, the photon exchange region ($0.4\text{-}0.7$~meV), the current dip is enhanced as $\kappa$ is increased, which is opposite to what is obtained in \fig{fig04}. This can be explained by looking at the characteristics of the first-excited state and its replicas in \fig{fig08}(e-g) and \fig{fig09}(e-g), in which the partial occupation and the partial current increase as $\kappa$ is increased. These characteristics are reflected in the behavior for the total current.

In \fig{fig08} the partial occupation of the DQD system as a function of the photon energy for (a) the ground state, (b) the first-, (c) the second-, (d) the third-photon replica of ground state, and (e) the first excited state, (f) the first-, (g) the second-photon replica of the first-excited state, for three values of the cavity-environment coupling $\kappa=10^{-5}$ (red), $10^{-4}$ (green), and $10^{-3}$ meV (blue) in case of $x$-polarized photon field. At low photon energy a constant occupation is seen for all states, also almost the same occupation obtained for all three values of the cavity-environment coupling. In the photon exchange region, $0.4\text{-}0.7$~meV, when the Rabi-resonances are expected to have a role, the ground and its photon replica states are populated, and the magnitude of occupation is reduced as $\kappa$ is increased.
The depopulation of the first-excited state and its photon replica states increases as $\kappa$ increases.
The main reason is that with increasing $\kappa$, the photon leakage of the cavity is enhanced leading to decrease of the Rabi-resonances and thus the Rabi-oscillation. The occupation of the ground state and its replicas originating from the Rabi-resonances is thus decreased, while the occupation of first-excited state and its replicas is enhanced for higher values of $\kappa$ in the intermediate photon energy range,  $0.4\text{-}0.7$~meV. This is because the first-excited state is located in the bias window and it is expected to have a higher occupation and current through the system even when the Rabi-resonances become weaker.

\fig{fig09} displays the partial current from the left lead to the DQD system as a function of photon energy for (a) the ground state, (b) the first-, (c) the second-, (d) the third-photon replica state of ground state, and (e) the first excited state, (f) the first-, (g) the second-photon replica state of the first-excited state, for three values of the cavity-environment coupling $\kappa=10^{-5}$ (red), $10^{-4}$ (green), and $10^{-3}$ meV (blue) in case of an $x$-polarized photon field. Again as the ground state is outside the bias window there is zero partial current through it, as for all other states it can be seen in the intermediate photon energy that the current is enhanced as the cavity-environment coupling is increased. In the high photon energy range the opposite behavior is seen for the first-excited state and its replica states. The highest value of the partial current is for the first-excited state \fig{fig08}(e), this is because of the presence of this state inside the bias window. It can be said that the first-excited state makes the highest contribution to the total current and thus the total current follows the characteristics of this state.

\section{Conclusion}\label{conclusion}

We have studied the transport properties of a DQD system connected to leads and coupled to a single mode photon cavity in the strong coupling regime using a quantum master equation appropriate for a many-level system in the steady state. The influence of the electron-photon and the cavity-environment coupling strengths on the transport properties for both the $x$ and the $y$-polarization of the photon field, is investigated. Tuning the frequency of the cavity photon field results in a noticeable change in the current through the DQD system. Due to the photon exchange of states contributing to the transport in the intermediate photon energy range caused by multiple Rabi we observe a current dip. The current dip is reduced as the cavity-environment coupling increased, while an increase of the electron-photon coupling strength leads to a decrease in the current throughout the total photon energy ranges investigated. In these calculations, due to the special symmetry of the two embedded dots, the polarization direction of the photon field does not affect their transport properties.

\section{Acknowledgment}
This work was financially supported by the University of Sulaimani and
the Research center of Komar University of Science and Technology.
The computations were performed on the IHPC facilities provided by the University of Iceland.



\end{document}